\documentclass[aps,prl,onecolumn,superscriptaddress,16pt]{revtex4}
\usepackage{amsmath}
\usepackage{amssymb}
\usepackage{graphicx}
\usepackage{footnote}
\usepackage{subfigure}
\usepackage{sidecap}
\usepackage{verbatim} 
\usepackage{microtype}
\usepackage{hyperref}
\usepackage{float}
\usepackage{bm}
\usepackage{setspace}
\begin{document}
\title{Anomalous Diffusion and Stress Relaxation in Surfactant Micelles}
\author{Subas Dhakal}
\email{sdhakalsnu@gmail.com}
\affiliation{Department of Biomedical and Chemical Engineering, Syracuse University, Syracuse, NY 13244}
\author{Radhakrishna Sureshkumar}
\email{rsureshk@syr.edu}
\affiliation{Department of Biomedical and Chemical Engineering, Syracuse University, Syracuse, NY 13244}
\affiliation{Department of Physics, Syracuse University, Syracuse, New York 13244, USA}
\date{\today}

\begin{abstract}

We present the first molecular dynamics study to probe the mechanisms of anomalous diffusion in cationic surfactant micelles in the presence of explicit salt and solvent-mediated interactions. Simulations show that when the counter ion density increases, saddle-shaped interfaces manifest leading to the formation of branched structures. In experiments, branched structures exhibit lower viscosity as compared to linear and wormlike micelles, presumably  due to stress relaxation arising from the sliding motion of branches along the main chain. Our simulations provide conclusive evidence and a mechanism of branch motion and stress relaxation in micellar fluids.   Further, depending upon the surfactant and salt concentrations, which in turn determine the microstructure, we observe normal, subdiffusive and superdiffusive motion of surfactants. Specifically, superdiffusive behavior is associated with branch sliding, breakage and recombination of micelle fragments as well as constraint release in entangled systems.\end{abstract}
\maketitle

\begin{figure}
\includegraphics[width=.40\textwidth]{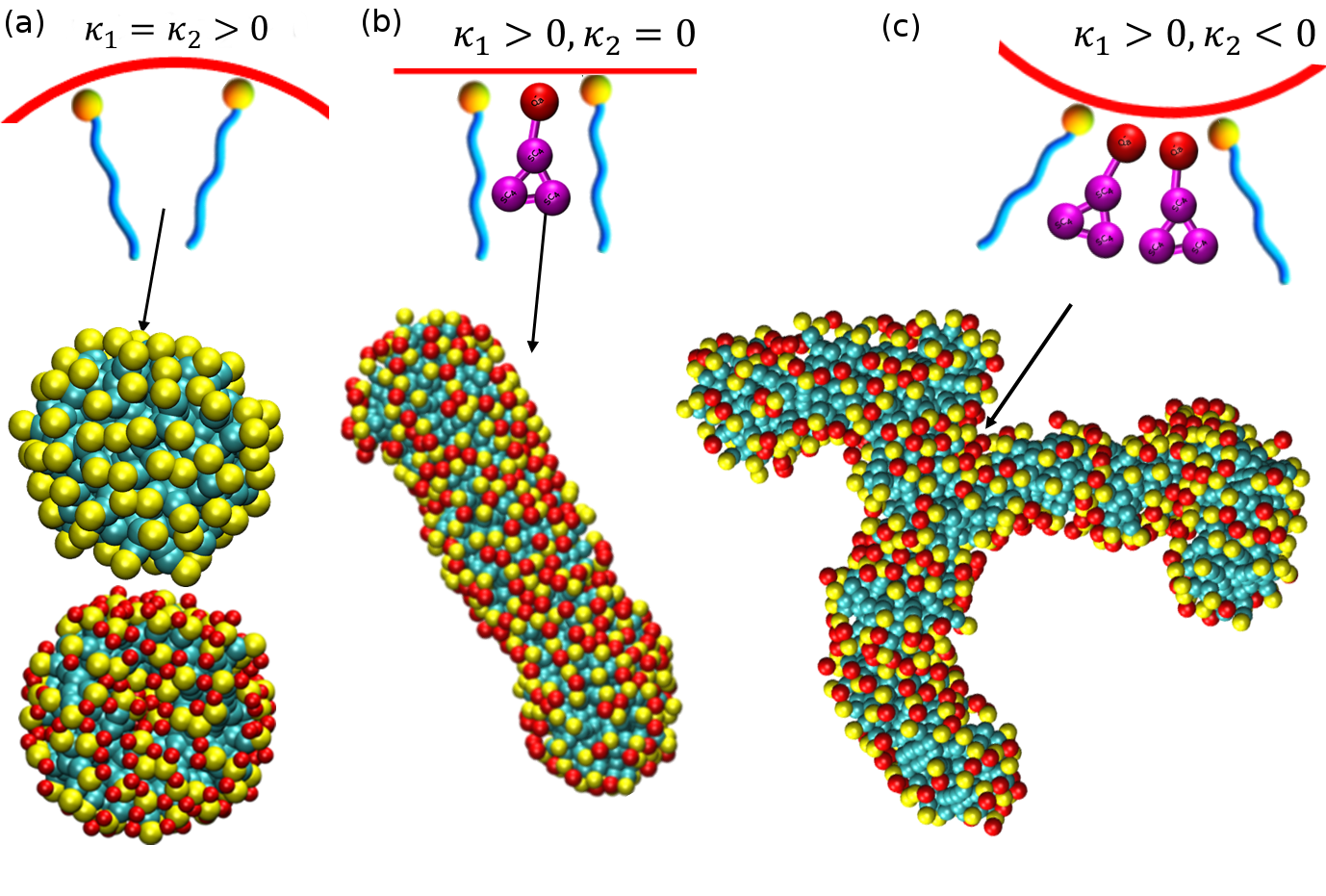}
\caption{\label{figure1}(Color online) Hydrophobic-water interfaces of micelles at c$_D$=0.2 M with different Gaussian curvature ($K$): (a) spherical, $K > 0$ (R=0 (top), R=1.2 (bottom)), (b) cylindrical (R=1.2), $K=0$, (c) hyperbolic (R=1.2), $K < 0$. Changes in the shape and curvature are induced by the condensation of the counter ions as depicted in the top panel.}
\end{figure}

Over the past decades, the structure, dynamics and mechanical properties of self-assembled aggregates of cationic surfactants have been studied extensively \cite{Debye51, Danino95, Lequeux96, CatesCandau90, Appell92, Sangwai11a, Abhi15, Rogers14, Subas15a, May97, Subas15b, Ott90, Ganapathy07, Gonzalez06, Angelico98, Lin96, Silvander96,Candau01,Sachsenheimer14,Ziserman09}. Self-assembly of cationic surfactant solutions can be controlled by manipulating the solvent-mediated electrostatic interactions among the surfactant molecules by altering the counter ion concentration. A rich variety of fluctuating micelle morphologies can be thus formed such as spheres, cylinders, wormlike chains as well as branched and loopy structures \cite{Lin96, Rogers14, Subas15a}. Aggregate shape, which depends on the solution temperature as well as the concentrations of the surfactant and the counter ion, has a profound effect on the rheological properties of micelle solutions.  Both experiments and recent simulations \cite{Lin96, Rogers14, Subas15a} have shown the existence of branched structures in ionic surfactant solutions. It has been hypothesized that, unlike in branched polymers, thermal fluctuations can cause micelle branches to slide along the contour of the main chain and thereby provide an additional mechanism of stress relaxation \cite{Rogers14,Candau01, Sachsenheimer14, Ziserman09} . Electrostatic screening by counter ions is known to promote branch formation and an accompanying reduction in the solution viscosity \cite{Sachsenheimer14,Subas15a}. Further, branching tends to suppress shear banding instabilities in wormlike micelle solutions \cite{Rogers14}. Despite the extensive literature on branched micelle systems, two major questions regarding their structure and dynamics remain unanswered:

\emph{i. Mechanisms and energetics of branching:} Branched structures are formed in a cationic surfactant solution as the salinity is increased \cite{Lin96, Rogers14, Sachsenheimer14,Subas15a}. However, the physical mechanisms underlying their formation are unclear. Second, direct molecular-level visualizations have not yet been performed to study the energetics underlying branched structures. Third, a direct evidence of branch sliding and its influence on stress relaxation are lacking in the literature.  Quantifying such putative driven motion of micelle branches poses a great challenge to experiments and molecular simulations. In this work, we have addressed the abovementioned questions via a coarse-grained molecular dynamics (CGMD) simulation of cationic surfactant micelle solutions of cetyl-trimethyl-ammonium-chloride (CTAC), in the presence of a binding organic salt sodium salicylate (NaSal), a widely used system in experimental studies \cite{Lequeux96, CatesCandau90, Appell92, Sachsenheimer14}. 
\begin{figure*}
\includegraphics[width=.90\textwidth]{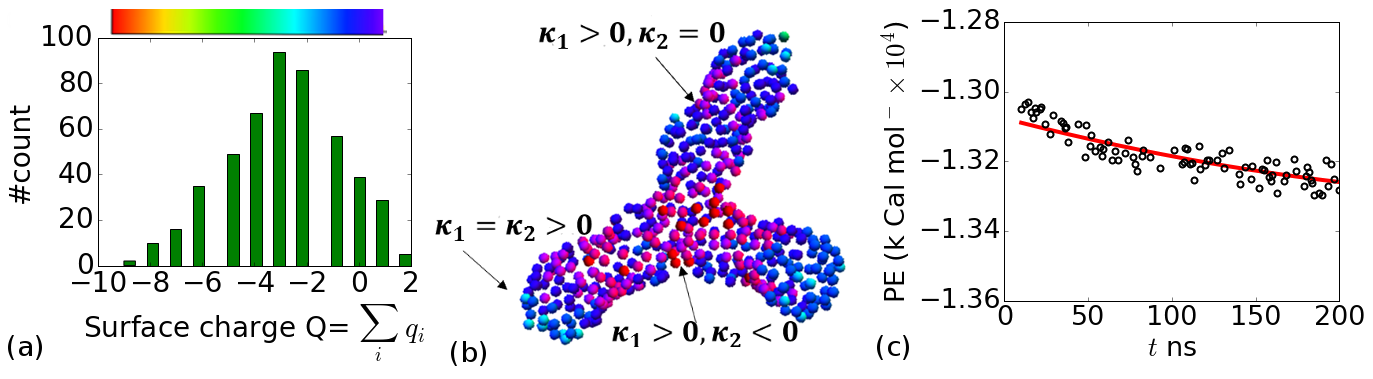}
\caption{\label{figure2}(Color online) (a) Distribution of net charge $Q$ over the micelle surface. The color map corresponds to the $Q$ values of the abscissa. (b) Micelle surface color-coded with $Q$ values. (c) The potential energy of a branched micelle as a function of time associated with the sliding motion.} 
\end{figure*}

\emph{ii. Anomalous diffusion:} Anomalous diffusion is ubiquitous in many phases of soft condensed matter. Subdiffusive behavior is encountered in crowded systems due to topological constraints, e.g. the reptative motion of polymer chains in an entangled polymer melt. On the other hand, superdiffusion is less common. Examples include molecular diffusion in glass forming liquids \cite{Winter12}, micro-scale particle diffusion in bacterial suspensions \cite{Wu14} and spin transport in Heisenberg quantum magnets \cite{Hild14}.  Wormlike micelle solutions exist in a dynamic equilibrium determined by breakage and recombination events and are inherently polydisperse. Such dynamics and entanglement effects can cause large variations in the mobility of individual surfactants within the system \cite{Ganapathy07, Ott90}.  For instance, the lifetime of trapping due to entanglements would depend on the micelle length and hence the motion of individual surfactant molecules could vary from one aggregate to the other. In fact, Ott \emph{et al.} \cite{Ott90} observed anomalous self-diffusion of surfactants in a solution of cetyl trimethyl bromide by measuring the diffusion of fluorescent probes using the fringe-pattern photobleaching (FRAP) technique. Specifically, the mean-square displacement (MSD) of the probe molecules was found to scale as $\langle \left[\vec{r}(t)-\vec{r}(0)\right]^2 \rangle\sim t^\beta$, with $\beta\neq 1$. For times shorter than the reptation time ($t\le \tau_r$), superdiffusive behavior was often observed while subdiffusive behavior was more common for $\tau >> \tau_r$. On the other hand, normal $\beta = 1$ and subdiffusive $\beta \le 1$ behaviors were observed in dilute solutions for long times. These authors attributed the superdiffusive motion to a \emph{Levy flight} caused by the reptation of shorter micelles. In this \emph{Letter}, we quantify surfactant motion at various salt ($c_S$) and surfactant ($c_D$) concentrations to understand the mechanisms of anomalous diffusion in CTAC-NaSal solutions. 

CGMD simulations of a model \cite{Subas15a, Sangwai11a, Abhi15} of CTAC-NaSal solution in water are conducted using the LAMMPS \cite{Plimpton95} molecular dynamics package.  This model utilizes the MARTINI force field \cite{Marrink07} for representing the interactions among the CG beads. A Particle-Particle Particle-Mesh solver is used to compute long-range electrostatic interactions.  The equation of motion is integrated for a constant NVT ensemble with the temperature controlled via a Nose-Hoover thermostat with a time step $\Delta t = 15 fs$. 
\begin{figure}
\includegraphics[width=.40\textwidth]{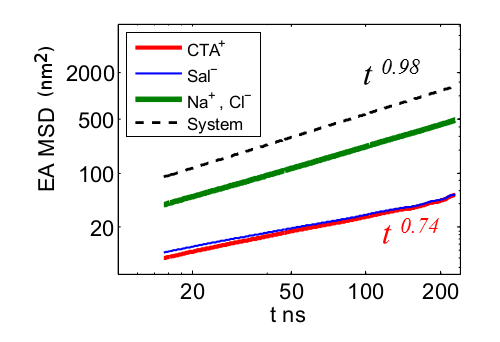}
\caption{\label{figure3}(Color online) EA MSD for different molecules/ions in a solution for $c_D=0.1$ M, and $R=0.2$. }
\end{figure}
\begin{figure*}
\includegraphics[width=.70\textwidth]{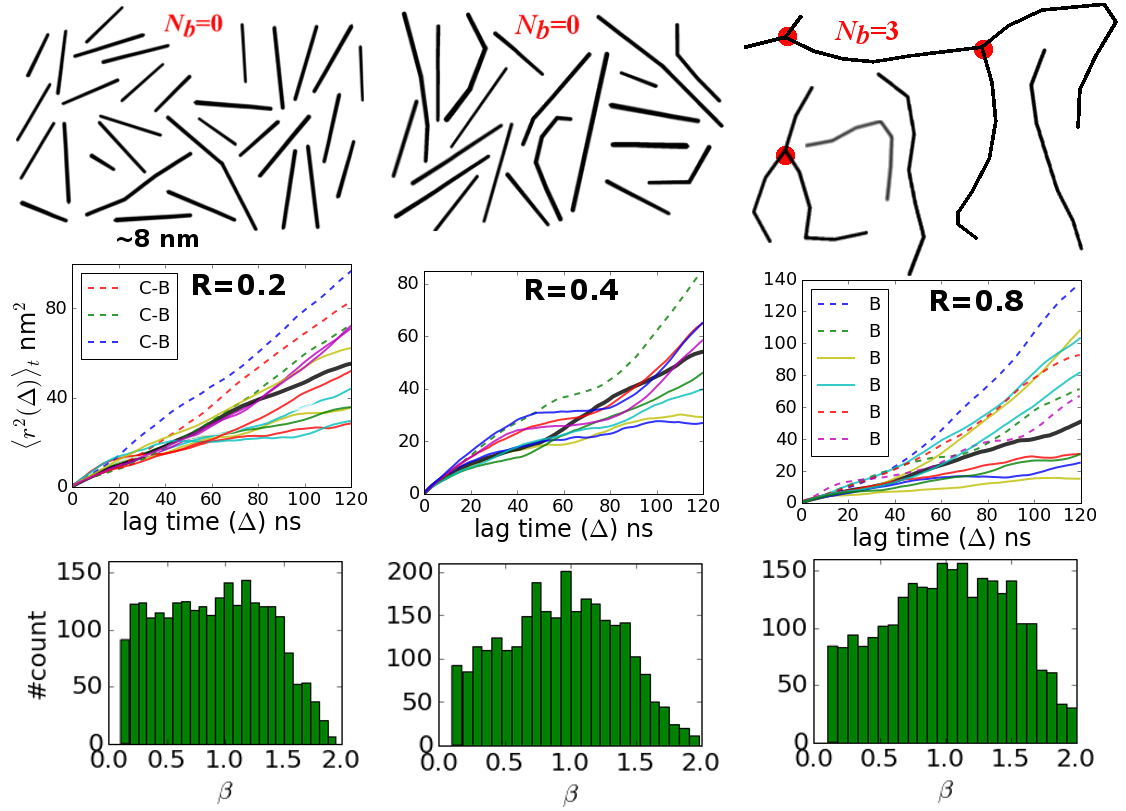}
\caption{\label{figure4} (Color online) First row: Microstructures for $R=0.2, 0.4$ and $0.8$ at $c_D=0.1$ M. Second row: TA MSD of representative surfactants showing normal, subdiffusive and superdiffusive motion. (C-B: surfactant belonging to a micelle that either combines with another micelle or breaks apart into two micelles, B: surfactant belonging to a branched micelle). Thick solid lines (black) represent normal diffusion, i.e., $\langle r^2(\Delta) \rangle \sim t$. Third Row: Distribution of the diffusion exponent $\beta$. }
\end{figure*}

We present typical examples of micelle-water interfaces in Fig. \ref{figure1} formed at different salt and surfactant concentrations. Specifically, surfaces with positive, zero and negative Gaussian curvature $K$ corresponding to spherical, cylindrical and branched micelles with saddle-shaped junctions are observed. To understand the energetics of branch formation, we analyzed the counter ion distribution in a typical \emph{Y-shaped} branched structure shown in Fig. \ref{figure1} c. We divided the micelle surface into patches of radius $\sim$ 1nm and calculated the effective charge $Q$ by summing over the partial charges within the patch. We show the distribution of $Q$ over the micelle surface in Fig. \ref{figure2} a. Surfactant head groups color-coded with the $Q$ value on the surrounding patch is shown in Fig. \ref{figure2} b.  Three distinct regions can be observed, namely spherical caps (blue), cylindrical regions (magenta) and a saddle-shaped junction (red).  At the saddle junction, surfactant orientations are aligned closely with the surface normal as shown schematically in Fig. \ref{figure1} c. This enhances the electrostatic repulsion between the surfactant head groups and makes the saddle geometry energetically unfavorable. Further, as inferred from Fig. \ref{figure2} a-b, a larger number of Sal$^-$ ions interdigitate among the surfactant molecules within the branched portion as compared to those in spherical and cylindrical regions. Hence, the additional electrostatic repulsion facilitated by this increase in counter ion condensation at the micelle-water interface compounds the unfavorable curvature energy of the saddles. Consequently, the branches are inherently unstable and incessantly move along the micelle contour as visualized in Movie S1 (SI) for $c_D = 0.3$ and $R = 0.8$. Quantitatively, the branch shown in the movie moves approximately $\sim 10$ nm in 50 ns. We tracked the potential energy (PE), which is the sum of the non-bonded interaction energies among all the CG beads within the branched structure, associated with this motion during which counter ions continually dissociate from and associate with the structure. As shown in Fig. \ref{figure2} c, the net effect of such local concentration fluctuations is a reduction in the PE along the direction of sliding. We note that fluctuations in counter-ion concentration could render a given sliding path energetically unfavorable, resulting in a reversal in the direction of branch motion. The persistence time of the sliding motion along a given direction is on the order of 10 ns.   

We conduct MD simulations at $c_D=0.1, 0.2, 0.3$ M for various $R= c_S/c_D=0.2, 0.4, 0.6, 0.8$ values to understand the mechanisms of branch formation. The size of our simulation box is approximately 35 nm in each dimension and consists of approximately $5 \times 10^5$ CG beads including water, surfactants, salts, and ions (see details in Ref. \cite{Subas15a}). As shown in Figure S1, numerous pathways of branch formation were identified including those involving multiple linear and/or branched micelles resulting in a variety of micelle morphologies.

\begin{figure*}
\includegraphics[width=.80\textwidth]{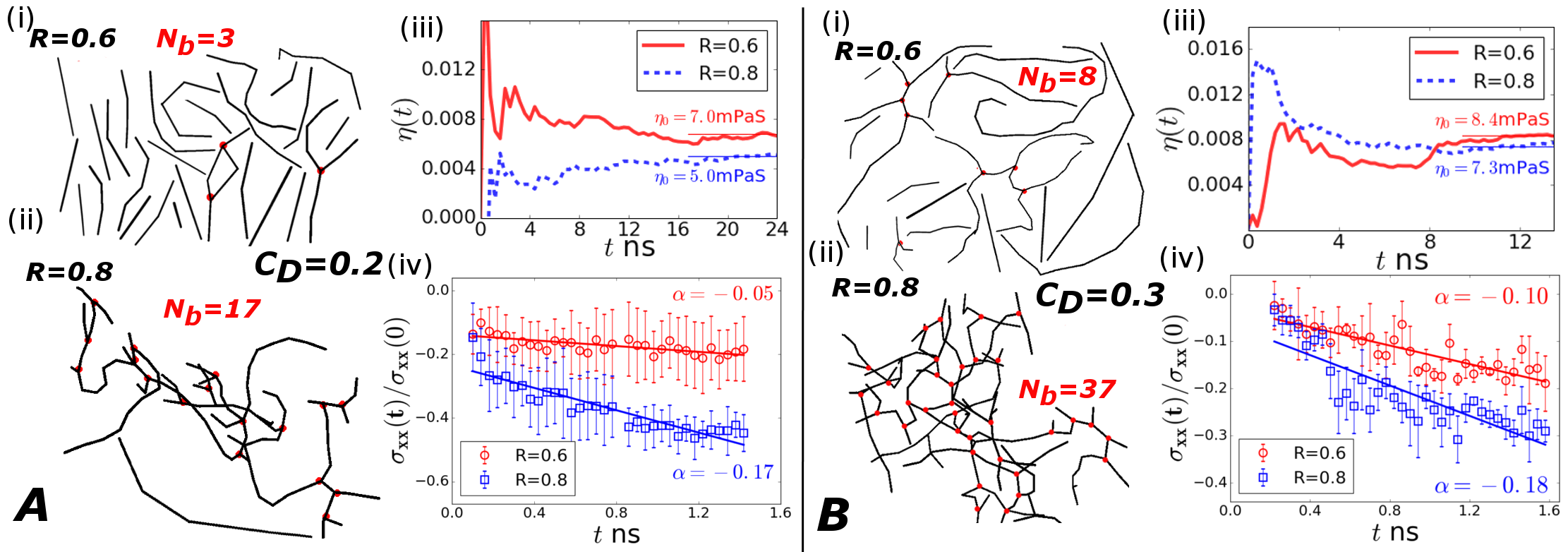}
\caption{\label{figure5}(Color online) ) Stress relaxation simulations. (A) $c_D=0.2$ M. (B) $c_D=0.3$ M. (i-ii) Microstructures at different $R$ values. (iii) Viscosity $\eta$ as a function of time for different concentrations. (iv) Time evolution of shear stress at different salt/surfactant concentrations after the cessation of an imposed extensional deformation (semi-log $y$ plot).}
\end{figure*}

Ensemble-averaged (EA) mean square displacement (MSD) of different molecules/ions for $c_D=0.1$ M and $R=0.2$ is shown in Fig.\ref{figure3}. Diffusion of the entire system is approximately Gaussian ($\beta \sim 0.98$) whereas those of the surfactants and Sal$^-$ counter ions are subdiffusive ($\beta \sim 0.74$) which is consistent with the notion that they constitute the self-assembled micellar structure. However,   persistent branch motion as well as breakage and recombination of micelle fragments could result in marked differences in the dynamics at the single molecule level. In order to probe such dynamic heterogeneity, we analyze the time-averaged mean square displacement (TA MSD) of the individual surfactants defined as 
\begin{equation}
\langle r^2\left(\Delta\right)\rangle_t = \frac{1}{T-\Delta}\int_0^{T-\Delta} \left[\bf{r}\left(t+\Delta\right)-\bf{r}(t)\right]^2 dt,
\end{equation}
where $\bf{r}(x,y,z)$ is the position vector of the surfactant center of mass, $\Delta$ and $T$ are the lag time and total measurement time respectively. In all calculations reported here, $T>200 ns$. We show TA MSD of representative surfactant molecules at various surfactant concentrations and $R=0.2, 0.4,$ and $0.8$ in Fig.\ref{figure4} along with the topological representations of the corresponding microstructures (top row) constructed by juxtaposing the $3D$ micelle contour onto a $2D$ plane. The branch points are denoted by red dots.  TA MSD plots show wide variation in surfactant dynamics with a broad distribution in the exponent, i.e., $0 \le \beta \le 2$. Subdiffusive motion results from the trapping of micelles in locally crowded environments and persists at all concentrations. On the other hand, the mechanism underlying superdiffusive motion is concentration dependent. Superdiffusive motion in a dilute solution (Fig. \ref{figure4} a) arises primarily due to transient combination and breakage of micelles as illustrated in animation S2. At intermediate concentrations (Fig. \ref{figure4} b), micelles are relatively longer, and reptation of micelle chains dominate the stress relaxation process. However, recombination and breakage still occur among shorter micelles. TA MSD of a surfactant molecule that belongs to a shorter micelle entrapped in entanglements (signified by the plateau regions in Fig. \ref{figure4} b) exhibits superdiffusive behavior when the topological constraints are released and the micelle escapes the “trap” (see animation S3).  Branched structures form upon further increasing $R$ to 0.8. Consequently, we often observe superdiffusive behavior in surfactants belonging to one of the branches of a micelle as shown in Fig. \ref{figure4} c. This stems from the motion of micelle branches as illustrated in animation S4. Diffusion in a micellar fluid occurs by solvent-mediated exchange of free surfactants, recombination and breakage of micelles, sliding motion of micelle branches, and reptation. Reptation inhibits free diffusion of molecules while micelle combination and sliding motion of branches enhance diffusion. Further, the distributions of the exponent $\beta$ shown on the bottom row of Fig. \ref{figure4} show that the number of surfactants exhibiting superdiffusive behavior increases with increasing $R$. This is consistent with the notion that branch motion as well as combination and breakage of micelles   synergistically boost surfactant diffusion at higher $R$, while combination and breakage is the primary mechanism of superdiffusion at lower $R$.

To investigate the influence of micelle branching on stress relaxation, we study the transient response of micellar solutions at various surfactant concentrations: $c_D=0.2$ M, and $0.3$ M for $R=0.6$ and $0.8$ after the cessation of a uniaxial extensional flow. We consider microstructures shown in Fig. \ref{figure5}(A-B) i-ii which clearly indicate an increase in the number branch points $N_b$ with increasing $R$. A large $N_b$ should then decrease the solution viscosity \cite{Subas15b} as a consequence of the enhanced diffusion of branch points. Therefore, one would expect a direct dependence of stress relaxation and molecular diffusion on the degree of micelle branching. To test this hypothesis, we compute the viscosity $\eta (t)$ for solutions with varying degree of branching (Fig. \ref{figure5} (A-B) iii) from the stress-auto correlation function $\eta (t)=\frac{V}{k_B T}\sum_{i,j} \int_0^\infty dt \langle p_{ij}(t).p_{ij}(0)\rangle$ with $ i\neq j=x,y,z$, where $V$ is the system volume, $k_B$ is the Boltzmann constant and $T$ is the temperature. As shown in Fig. \ref{figure5} iii, the zero-shear viscosity $\eta_0$, which is the plateau in $\eta (t)$, decreases with increasing $N_b$ in accord with previous studies \cite{Subas15a, Sachsenheimer14, Rogers14}.  To understand the underlying mechanism for this reduction in viscosity and the role of branching on stress relaxation, we study how an externally imposed stress relax in such systems. Towards this end, we deform the microstructures under a uniaxial tensile strain at a constant strain rate in the $x$ direction with zero pressure in $y$ and $z$. Such a flow stretches the microstructure in the $x$ direction while shrinks it in the other two directions \cite {Subas15b}. In all of the simulations, the applied deformation is such that  the equilibrium structures are appreciably stretched but yet retained their original topologies. In Fig. \ref{figure5} (A-B) iv, we plot the residual stress $\sigma_{xx} (t)$ after the cessation of extensional flow. In all cases, stress decreases exponentially $\sigma_{xx} (t)\sim \sigma_{xx}(0) \exp^{-\alpha t}$. Clearly, stress relaxation is faster in a solution with a higher degree of branching. Closer inspection of the individual micelles in the simulations reveals that the branches indeed slide along the main chain as conjectured by Appell \emph{et al.} \cite{Appell92}. This study provides clear evidence that such sliding motion helps heal stresses imposed on the system.

In conclusion, we have explored the dynamics of surfactant molecules in micellar solutions with diverse morphologies. Simulations have revealed multiple pathways of branch formation and provided direct visualizations of the sliding motion of micellar branches. We have also identified the energetic driving forces that induce incessant branch motion arising from an excess counter ion condensation at the branch points. Overall, diffusion of surfactant molecules within a micelle solution is strikingly heterogeneous as signified by the existence of a broad range of subdiffusive, normal and superdiffusive space-time trajectories. Superdiffusion of surfactant molecules can arise from three mechanisms: sliding motion of branches along the micelle contour, breakage and recombination of micelle fragments, and ballistic motion of short micelles that escape from topological constraints. The latter mechanism is supported by FRAP measurements \cite{Ott90}. Finally, the CGMD simulations have provided clear evidence to validate a hypothesis that was put forward almost three decades ago that an increase in branch density facilitates faster stress relaxation in micellar fluids. 

\acknowledgments
This work used the computational resources provided by Extreme Science and Engineering Discovery Environment (XSEDE), which is supported by National Science Foundation grant number OCI-1053575. The authors acknowledged financial support by National Science Foundation under Grants 1049489 and 1049454.   

\end{document}